\documentclass[aps,pra,twocolumn,showpacs,floatfix]{revtex4-1}

\usepackage[T1]{fontenc}
\usepackage{amsmath}
\usepackage{txfonts}
\DeclareMathAlphabet{\mathbit}{OT1}{txr}{b}{it}
\usepackage{microtype}
\usepackage{graphicx}
\graphicspath{{figures/}}
\usepackage{color}
\usepackage[breaklinks=true,
pdfauthor={Michael Klaiber, Karen Z. Hatsagortsyan},
pdftitle={Laser-driven relativistic tunneling from $p$-states}]{hyperref}
\usepackage{braket}

\begin{document}

\title{Laser-driven relativistic tunneling from $p$-states}

\author{Michael Klaiber}
\author{Karen Z. Hatsagortsyan}
\affiliation{Max-Planck-Institut f{\"u}r Kernphysik, Saupfercheckweg~1,
  69117~Heidelberg, Germany}
\date{\today}

\begin{abstract}
The tunneling ionization of an electron from a $p$-state in a highly charged ion in the relativistic regime is investigated in a linearly polarized strong laser field. In contrast to the case of an $s$-state, the tunneling ionization from the $p$-state is spin asymmetric. We have singled out two reasons for the spin asymmetry: first, the difference of the electron energy Zeeman splitting in the bound state and during tunneling, and second, the relativistic momentum shift along the laser propagation direction during the under-the barrier motion. Due to the latter, those states are predominantly   ionized  where the electron rotation is opposite to the electron relativistic shift  during the under-the-barrier motion.  We have investigated the dependence of the ionization rate on the laser intensity for different projections of the total angular momentum and identified the intensity parameter which governs this behaviour. The significant change of the ionization rate is originated from the different precession dynamics of the total angular momentum in the bound state at high and low intensities.

\end{abstract}

\pacs{32.80.Rm, 31.30.J-}

\maketitle

\section{Introduction}

Recently, significant experimental efforts are invested for investigation of relativistic regimes of strong field ionization \cite{DiChiara_2008,Palaniyappan_2008,DiChiara_2010,Ekanayake_2013} which fostered accordingly the development of theory \cite{RMP_2012}. In particular, specific signatures of the relativistic under-the-barrier dynamics in the photoelectron momentum distribution have been pointed out recently in \cite{Klaiber_2013c,Yakaboylu_2013,Yakaboylu_2014b} and subtle spin effects in laser fields are explored in \cite{Ahrens_2012,Ahrens_2013,Skoromnik_2013,Wen_2014,Ahrens_2014,Bauke_2014a,Bauke_2014b}. 
During the relativistic laser-atom interaction spin effects were shown to appear in the laser-driven bound electron dynamics \cite{Walser_2002} and, in particular, in the radiation of high-order harmonics \cite{Hu_1999,Walser_2001}. Spin effects arise also during tunneling ionization \cite{Faisal_2004,Klaiber_2014}. The spin effects in nonsequential double ionization of helium were considered in \cite{Bhattacharyya_2007,Bhattacharyya_2011}. It appeared that during the relativistic tunneling ionization from a ground state of an hydrogenlike ion spin asymmetry is negligible, however, the spin flip is possible when using relativistic laser intensities of order of $10^{22}$ W/cm$^2$  and highly charged ions, with the charge state of order of $Z\sim 30$. A question arises if spin asymmetry can exist for ionization of non-spherical symmetric states?

In the nonrelativistic regime the strong field ionization rate in the tunneling regime is calculated in the Perelomov-Popov-Terent'ev (PPT) theory \cite{Perelomov_1966a,Perelomov_1966b,Perelomov_1967a,Ammosov_1986,Popov_2004} for any value of the angular momentum $l$ and the magnetic quantum number $m$. The laser pulse effect in ionization of excited states of an hydrogen atom in the nonrelativistic regime is considered in \cite{Zon_2005}. The excited $p$-state of He$^+$ has been proposed to employ for control of the polarization of isolated attosecond pulses \cite{Liu_2012}. In this context, the peculiarities of strong field ionization, recollision and high-order harmonic generation from antisymmetric molecular orbitals are also known \cite{Lein_2003,Kjeldsen_2003,Fischer_2006}. Note that the ionization from $m_l\neq 0$ states is an essential ingredient in the dynamics of multiple  ionization of the atomic target in ultrastrong laser fields \cite{McNaught_1997,Moore_1999,Yamakawa_2004,Gubbini_2005,Taieb_2001}.

Recently, the interest to the strong field ionization of an electron from a $p$-state has been renewed in connection with the non-adiabatic ionization in a circularly polarized laser field \cite{Barth_2011,Barth_2013a,Herath_2012}. It turned out that in the non-adiabatic regime, when the Keldysh parameter $\gamma$ \cite{Keldysh_1965} is not small, the electron in the bound state rotating opposite to the field rotation ($m<0$) is ionized easier than in the co-rotating case. Moreover, a spin polarization effect is found in \cite{Barth_2013b} due to the interplay of the electron-core entanglement and the sensitivity of ionization in a circularly polarized field to the magnetic quantum number $m_l$.

In this paper we consider tunnel-ionization in the relativistic regime from an exited $p$-state of an hydrogenlike ion induced by a strong linearly polarized laser field. The main concern is to investigate the dependence of the tunneling probability on the magnetic and spin quantum numbers in the relativistic regime and to find conditions when a  large spin asymmetry can exist. We will show that in the relativistic regime even in the adiabatic case $\gamma \ll 1$ one can observe the dependence of the ionization probability on the magnetic quantum number similar to the nonrelativistic non-adiabatic regime. 
Those bound states are predominantly  ionized  where the electron rotation in the bound state is opposite to the electron relativistic shift along the propagation direction during the under-the-barrier motion.

The tunneling ionization rates from the excited $p$-states of a hydrogenlike ion are calculated. For more convenience the relative ionization rates of the $p$-state with respect to the $s$-state, rather than the absolute ionization rates, are presented. For this reason we first calculate the tunneling ionization rate from the excited $2s$-state of the hydrogenlike ion. We employ a Coulomb-corrected relativistic Strong Field Approximation (SFA) developed in \cite{Klaiber_2013a,Klaiber_2013b} for the description of the ionization dynamics in the relativistic regime.

\section{Calculation of the ionization rate}

The ionization differential rate is expressed via the transition matrix element 
\begin{eqnarray}
  \frac{dw}{d^3\mathbf{p}}=\frac{\omega}{2\pi}|M|^2,
  \label{dWdp}
  \end{eqnarray}
with the laser frequency $\omega$. The matrix element $M$ in the Coulomb corrected SFA  reads \cite{Klaiber_2013b}: 
\begin{eqnarray}
  M=\int dt d^3\mathbf{r}\, \psi^V_C(\mathbf{r},t) \,\mathbf{r}\cdot\mathbf{E}(\eta)\,\tilde{\psi}_{i}(\mathbf{r},t),
  \label{M}
\end{eqnarray}
where the final state $\psi^V_C$ is the Eikonal-Coulomb-Volkov-state \cite{Klaiber_2013b}, i.e., the wave function in the eikonal approximation for the electron in continuum under the action of the laser and Coulomb field of the atomic core; $\textbf{E}(\eta)=E_0 \cos(\omega\eta)$ is the laser field with the phase $\eta=t-z/c$, $c$ is the speed of light, and $\tilde{\psi}_{i}$ is the dressed initial bound state which is the solution of the following Schr\"odinger equation \cite{Klaiber_2013b}
\begin{eqnarray}
  i\partial_t\tilde{\psi}_{ i}=H_B\tilde{\psi}_{ i},
  \label{dressed_eq}
\end{eqnarray}
with the dressed bound state Hamiltonian
\begin{eqnarray}
  H_B=c{\boldsymbol \alpha}\cdot\left[\mathbf{p}-\hat{\mathbf{k}}\,\left(\mathbf{r}\cdot\mathbf{E}(\eta)\right)\right]+\beta c^2+V(\textbf{r}),
  \label{dressedH}
\end{eqnarray}
where $\hat{\mathbf{k}}$ is the unit vector in the laser propagation direction, $V(\textbf{r})$ is the potential of the ionic core, and ${\boldsymbol \alpha},\beta$ are the Dirac matrices.

First of all we calculate the dressed bound states. In the case of a $2s$-state the approximate solution of Eq. (\ref{dressed_eq}), taking into account only transitions between the states of the fine structure \cite{Klaiber_2013b}, is
\begin{eqnarray}
  \tilde{\psi}^{(2s)}_{ j}(\mathbf{r})=\psi^{(2s)}_j(\mathbf{r})\exp[ijA(\eta)/2c],
  \label{psi_2s}
\end{eqnarray}
where $\psi^{(2s)}_j(\mathbf{r})$ is the relativistic wave function of the initial $2s$-state of the electron in the highly charged hydrogenlike ion  \cite{Bethe_1957}:
\begin{eqnarray}
  \psi^{(2s)}_j(\mathbf{r})&=&\left(\left(1 - \sqrt{2I_p} r\right)\chi_j, i \sqrt{\frac{I_p}{2c^2}}\left(2- \sqrt{2I_p} r\right) \frac{\boldsymbol{\sigma}\cdot\mathbf{r}}{r}\chi_j\right)\nonumber\\
  &&\times\frac{\exp[-\sqrt{2I_p}r](2I_p)^{3/4}}{\sqrt{\pi}},
  \label{psi_i}
\end{eqnarray}
with the quantum number of the total angular momentum projection $j=(+,-)$, the spinors $\chi_+=(1,0)$,  $\chi_-=(0,1)$, the ionization energy $I_p$ and the Pauli matrices $\boldsymbol{\sigma}$ [the electron total energy in the bound state is $c^2-I_p$, and the ionization energy $I_p$ is related to the nuclear charge as $Z=2\sqrt{2I_p}$]. According to Eq. (\ref{psi_2s}) the $2s$-state experiences a Zeeman-splitting with an energy of $\varepsilon_J^{(b)}=-j\partial_t A/2c=g_S\mathbf{S}\cdot\mathbf{B}/2c$, where $S_B=\pm1/2$, $g_S=2$ and $A(\eta)=-E_0/\omega\sin(\omega\eta)$ .

Since the typical coordinate where the electron starts to leave the ion is $r\sim\sqrt{E_a/E_0}/\sqrt{2I_p}$ \cite{Klaiber_2013b}, with the atomic field strength $E_a=(2I_p)^{3/2}$ and $E_0/E_a\ll 1$ in the tunneling regime, the wave function of the initial state of Eq. (\ref{psi_i}) can be approximated
\begin{eqnarray}
  \psi^{(2s)}_j(\mathbf{r})&=&-\left(\chi_j, i \sqrt{\frac{I_p}{2c^2}}  \frac{\boldsymbol{\sigma}\cdot\mathbf{r}}{r}\chi_j\right)\frac{r\exp[-\sqrt{2I_p}r](2I_p)^{5/4}}{\sqrt{\pi}}.
\end{eqnarray}

The differential ionization rate from the $2s$-state is calculated via Eqs.~(\ref{dWdp})  and (\ref{M}). We have evaluated the differential rate for a given $p_E$ at the rate local maximum which is achieved at the momentum parabola   \cite{Klaiber_2013b} 
\begin{eqnarray}
p_k=\frac{I_p}{3c}+\frac{p_E^2}{2c}\left(1+\frac{I_p}{3c^2}\right) \,\,\,\,\,\,\,\,\,p_B=0,
\label{parabola}
\end{eqnarray}
where $p_E$, $p_B$ and $p_k$ are the momentum components along the laser electric field, the magnetic field and the propagation directions, respectively. On the mentioned momentum parabola the rate reads:
\begin{eqnarray}
  \frac{dw^{(2s)}_{\pm}}{d^3\mathbf{p}}=w_0\frac{12 c^4+p_{\text{E}}^2 \left(6 c^2-11 I_p\right)-18 c^2 I_p}{3 \left(2 c^2+p_{\text{E}}^2\right){}^2}\exp[-i S(p_E,\eta_s)],\nonumber\\
\end{eqnarray}
where the prefactor is $w_0\equiv 1024 \pi (2I_p)^{15/2}/E(\eta_s)^2$, $\pm$ refers to $m_j=\pm 1/2$ (in the case of the $s$-state the total angular moment is $J=1/2$). The time variable is changed to the phase variable $\eta$ in Eq.~(\ref{M}) and the $\eta$-integral is calculated with the saddle point method; $\eta_s$ is the saddle point value for the phase $\eta$~\cite{Klaiber_2013b}.
There exists no asymmetry between the ionization probabilities from the spin up and down states $2s_+$ ($m_j=1/2$) and $2s_-$ ($m_j=-1/2$), i.e., the ionization probabilities are equal.

For an intuitive understanding of the ionization spin asymmetry let us estimate the tunneling ionization probability via the WKB tunneling exponent:
\begin{eqnarray}
  \Gamma \sim \exp\left( -2\left|\int_0^{r_E^{(e)}} p_Edr_E\right| \right),
   \label{Gamma}
\end{eqnarray}
$r_E$ is the coordinate projection along the laser electric field,  $r_E^{(e)}$ is the tunnel exit coordinate. The electron momentum during the under-the-barrier-motion is complex and is derived form 
the energy conservation in the quasi-static tunneling picture:  
\begin{eqnarray}
 p_E^2/2+r_E E + \varepsilon_J^{(c)}= -I_p+\varepsilon_J^{(b)},
 \label{conserv}
\end{eqnarray}
where the left side of the equation is the  energy of the electron in the continuum during the tunneling and the right -- the energy in the bound state, with the Zeeman energy splitting in the continuum $\varepsilon_J^{(c)}$ and in the bound state  $\varepsilon_J^{(b)}$, respectively. From Eq. (\ref{conserv}) $p_E=i\sqrt{2(\tilde{\epsilon}+r_E E)}$, where $\tilde{\epsilon}=I_p-\Delta \varepsilon_J$ is the effective energy during tunneling including the angular momentum-magnetic field coupling, with $\Delta \varepsilon_J=\varepsilon_J^{(b)}-\varepsilon_J^{(c)} $. 
In the case of ionization from an $s$-state the Zeeman splitting has the same magnitude in the bound state and during tunneling $\varepsilon_J^{(b)}=\varepsilon_J^{(c)}=g_S\mathbf{S}\cdot\mathbf{B}/2c$. Therefore, the electron effective energy during the tunneling does not depend on the spin projection and, consequently, the tunneling probability is the same for both of the states $m_j=\pm 1/2$, explaining that there is  no spin asymmetry in this case.

For the $p$-states the total angular momentum can be $J=1/2$ or $3/2$. The wave function for the initial free bound state with $J=1/2$ reads \cite{Bethe_1957}
\begin{eqnarray}
  \psi^{(2p)}_{1/2+}(\mathbf{r})&=&\left\{-\frac{z}{r},-\frac{x+i y}{r},-\frac{i \left(\sqrt{2I_p} r-3\right)}{2 c r},0\right\}\frac{ (2I_p)^{5/4} r e^{-\sqrt{2I_p} r}}{\sqrt{3 \pi }},\nonumber\\
  \psi^{(2p)}_{1/2-}(\mathbf{r})&=&\left\{\frac{x-i y}{r},-\frac{z}{r},0,\frac{i \left(\sqrt{2I_p} r-3\right)}{2 c r}\right\}  \frac{(2I_p)^{5/4} r e^{-\sqrt{2I_p} r}}{\sqrt{3 \pi }}.\nonumber
\end{eqnarray}
They can be approximated analogously like the $2s$-states, yielding
\begin{eqnarray}
  \psi^{(2p)}_{1/2+}(\mathbf{r})&=&\left\{-\frac{z}{r},-\frac{x+i y}{r},-\frac{i \sqrt{I_p}}{\sqrt{2} c},0\right\}
  \frac{(2I_p)^{5/4} r e^{-\sqrt{2I_p} r}}{\sqrt{3 \pi }},\label{wavefunction1}\\
 \psi^{(2p)}_{1/2-}(\mathbf{r})&=&\left\{\frac{x-i y}{r},-\frac{z}{r},0,\frac{i \sqrt{I_p}}{\sqrt{2} c}\right\}
 \frac{(2I_p)^{5/4} r e^{-\sqrt{2I_p} r}}{\sqrt{3 \pi }}.
\end{eqnarray}
As can be seen from the wave function above, they are either a linear combination of $m_l=0,m_s=1/2$ and $m_l=1,m_s=-1/2$ or $m_l=-1,m_s=1/2$ and $m_l=0,m_s=-1/2$ with equal weighting, where $m_l,\,m_s$ are the quantum numbers for the orbital moment and spin projections.
The initial states in the SFA-amplitude, which are the eigenstates of the dressed atomic Hamiltonian of Eq. (\ref{dressedH}), equal
\begin{eqnarray}
  \tilde{\psi}^{(2p)}_{1/2\pm}&=&\psi^{(2p)}_{1/2\pm}\exp[\pm iA/6c], 
 \end{eqnarray}
The Zeeman energy splitting
\begin{eqnarray}
\varepsilon_J^{(b)}=g_J\mathbf{J}\cdot\mathbf{B}/2c
\end{eqnarray}
is determined by the Lande-factor
\begin{eqnarray}
g_J=\frac{3}{2}+\frac{S(S+1)-L(L+1)}{2J(J+1)},
\end{eqnarray}
with $S=1/2$, $L=1$, $J=1/2$, then $g_J=2/3$.

In the case of the $p$-states with $J=3/2$, we approximate similarly the exact wave function with $m_j=3/2$ as
\begin{eqnarray}
\psi^{(2p)}_{3/2++}(\mathbf{r})&=&\frac{(2I_p)^{5/4} r e^{-\sqrt{2I_p} r}}{\sqrt{2 \pi }}\\
&&\times\left\{\frac{  (x+i y)}{r},0,\frac{i  \sqrt{2I_p} z (x+i y)}{2 c r^2},\frac{i   \sqrt{2I_p} (x+i y)^2}{2 c r^2}\right\},\nonumber
\end{eqnarray}
which is a state with $m_l=1,m_s=1/2$. The state with $m_j=1/2$ is approximated as follows
\begin{eqnarray}
\psi^{(2p)}_{3/2+}(\mathbf{r})&=&\frac{(2I_p)^{5/4} r e^{-\sqrt{2I_p} r}}{\sqrt{6 \pi }}\\
&&\times\left\{\frac{2z}{r},-\frac{x+i y}{  r},\frac{i \sqrt{2I_p} \left(\frac{6 z^2}{r^2}-2\right)}{4 c},\frac{3 i \sqrt{2I_p} z (x+i y)}{2 c r^2}\right\},\nonumber
  \end{eqnarray}
which is a linear combination of $m_l=0,m_s=1/2$ and $m_l=1,m_s=-1/2$ in the ratio 2:1. The state with $m_j=-1/2$ is
\begin{eqnarray}
\psi^{(2p)}_{3/2-}(\mathbf{r})&=&\frac{(2I_p)^{5/4} r e^{-\sqrt{2I_p} r}}{\sqrt{6 \pi }}\\
&&\times\left\{-\frac{x-i y}{ r},-\frac{2z}{r},-\frac{3 i \sqrt{2I_p} z (x-i y)}{2 c r^2},\frac{i \sqrt{2I_p} \left(\frac{6 z^2}{r^2}-2\right)}{4 c}\right\}\nonumber
\end{eqnarray}
which is  a linear combination of $m_l=0,m_s=-1/2$ and $m_l=-1,m_s=1/2$ in the ratio 2:1. And finally the state with $m_j=-3/2$ reads
\begin{eqnarray}
\psi^{(2p)}_{3/2--}(\mathbf{r})&=&\frac{(2I_p)^{5/4} r e^{-\sqrt{2I_p} r}}{\sqrt{2 \pi }}\label{wavefunction2}\\
&&\times\left\{0,\frac{ (x-i y)}{r},\frac{i \sqrt{2I_p} (x-i y)^2}{2 c r^2},-\frac{i \sqrt{2I_p} z (x-i y)}{2 c r^2}\right\}\nonumber
\end{eqnarray}
which is a state with $m_l=-1,m_s=-1/2$. The dressed states in the SFA-amplitude are calculated:
\begin{eqnarray}
  \tilde{\psi}^{(2p)}_{3/2\pm\pm}&=&\psi^{(2p)}_{3/2\pm\pm}\exp[\pm iA/c],\nonumber\\
  \tilde{\psi}^{(2p)}_{3/2\pm\,\,}&=&\psi^{(2p)}_{3/2\pm}\exp[\pm iA/3c]\label{dressed2P3/2}
 \end{eqnarray}
Again with the help of the Lande-factor the Zeeman-energy splitting can be given, with $S=1/2$, $J=3/2$, $L=1$ and $g_J=4/3$ in this case.

In next sections we consider three physically relevant possible choices of the quantization axis for the angular momentum and spin: (a) along the laser magnetic field direction
(b) along the laser propagation direction and (c) along the electric field direction which will be considered in next sections.

\section{Quantization axis  along the laser magnetic field direction}

\begin{figure}
  \begin{center}
    \includegraphics[width=0.45\textwidth]{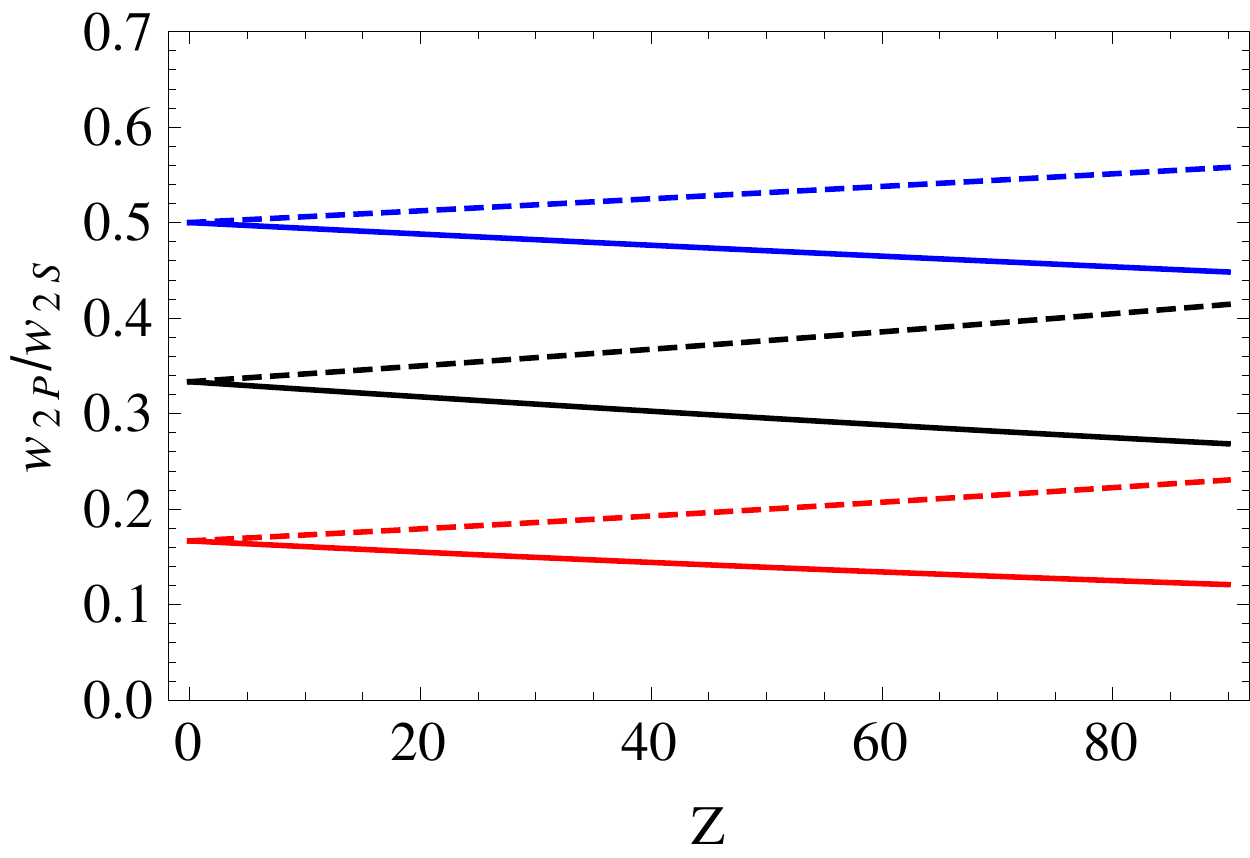}
    \caption{The relative as well as total ionization rates of $2p$-states $dw^{(2p)}/d^3\mathbf{p}$ with respect to the $2s$-state $dw^{(2s)}/d^3\mathbf{p}$ vs the hydrogenlike ion charge $Z$, when the angular momentum quantization axis is along the laser magnetic field. Blue (top lines):  $2p_{3/2++}\rightarrow\uparrow$ (dashed),  $2p_{3/2,--}\rightarrow\downarrow$ (solid); Red (bottom lines): $2p_{3/2+}\rightarrow\downarrow$ (dashed),   $2p_{3/2-}\rightarrow\uparrow$ (solid); Black (middle lines): $2p_{1/2+}\rightarrow\downarrow$ (dashed),  $2p_{1/2- }\rightarrow\uparrow$ (solid). The final electron spin is indicated by $\uparrow$ or $\downarrow$, the subscript $``\pm\pm"$ indicate $m_j=\pm 3/2$ and $``\pm"$ indicate $m_j=\pm 1/2$.}
    \label{B}
  \end{center}
\end{figure}

When the quantization axis is along the laser magnetic field direction, no spin flip can occur during ionization. Let us first consider the case of $J=1/2$.  We calculate the probability for the ionization from a $p$-state ($J=1/2$)  with the total angular momentum projection $m_j=\pm 1/2$. The corresponding relative differential probabilities with respect to the $2s$-state on the momentum parabola of Eq.~(\ref{parabola}) are
\begin{eqnarray}
  \frac{dw^{(2p)}_{1/2\pm}/d^3\mathbf{p}}{dw^{(2s)}/d^3\mathbf{p}}=  \frac{w^{(2p)}_{1/2\pm}}{w^{(2s)}}= \frac{1}{3} \pm \frac{2 \sqrt{2I_p/c^2}}{9}, \label{2P1/2+}
   \end{eqnarray}
where $I_p/c^2$ terms are neglected. The latter is illustrated in Fig. \ref{B}. There is a non-vanishing spin asymmetry with respect to tunneling ionization:
\begin{eqnarray}
  {\cal A}^{(2p)}_{1/2} =\left| \frac{dw^{(2p)}_{1/2+}/d^3\mathbf{p}-dw^{(2p)}_{1/2-}/d^3\mathbf{p}}{dw^{(2p)}_{1/2+}/d^3\mathbf{p}+dw^{(2p)}_{1/2-}/d^3\mathbf{p} }\right|\approx \frac{2}{3}\sqrt{\frac{2I_p}{c^2}}.
  \label{A}
\end{eqnarray}

Intuitively one can understand the ionization asymmetry between $2p_{1/2+}$ and $2p_{1/2-}$ states [the splitting of the middle line in Fig. \ref{B}] in the following way. In the nonrelativistic limit at this choice of the quantization axis, predominantly the states with $m_l=\pm 1$ are ionized [this corresponds to $m_l=0$ for more usual choice of the quantization axis along the electric field]. The state with $m_j=1/2$ are  the linear combination of $m_l=0,m_s=1/2$ and $m_l=1,m_s=-1/2$ states from which only the part of the bound state wave function with $m_l=1$ can be ionized which has here a weight of 1/2. Accordingly, the state with $m_j=-1/2$ are  the linear combination of $m_l=0,m_s=-1/2$ and $m_l=-1,m_s=1/2$ states from which only the part with $m_l=-1$ can be ionized. Therefore, the spin of the tunnelling electron in the states $m_j=\pm 1/2$ are opposite $m_s=\mp 1/2$. We can estimate the tunneling ionization probability via the WKB tunneling exponent Eq. (\ref{Gamma}):
\begin{eqnarray}
  \Gamma^{(2p)}_{1/2\pm}\sim  \exp\left( -\frac{4\sqrt{2}}{3}\frac{\tilde{\epsilon}^{3/2}}{E_0}  \right)\approx e^{ -\frac{2 }{3}\frac{E_a }{E_0}}\left(1+\frac{\Delta \varepsilon_J E_a}{I_pE_0} \right)
   \label{Gamma1}
\end{eqnarray}
where  $\Delta \varepsilon_J=g_J\mathbf{J}\cdot\mathbf{B}/2c-g_S\mathbf{S}\cdot\mathbf{B}/2c$
as the Zeeman energy splitting in the bound state is $\varepsilon_J^{(b)}=g_J \mathbf{J}\cdot\mathbf{B}/2c$, whereas during tunneling it is $\varepsilon_J^{(c)}=g_S\mathbf{S}\cdot\mathbf{B}/2c$. 
Then, $\Delta \varepsilon_J=2E_0/3c$, and according to Eq. (\ref{Gamma1})
\begin{eqnarray}
 \Gamma^{(2p)}_{1/2\pm} \sim e^{ -\frac{2 }{3}\frac{E_a }{E_0}}\left(1\pm \frac{4}{3}\sqrt{\frac{2I_p}{c^2}}\right), \end{eqnarray}
when $J=\pm 1/2$ and $S=\mp 1/2$.

Further, there is  a second reason for the asymmetry in the ionization probability. In a $p$-state  the electron rotates around the atomic core in the $k-E$-plane at $m_l=\pm 1$ (quantization axis is along the magnetic field). Since there is a shift due to the laser magnetic field in $k$-direction \cite{Klaiber_2013c,Yakaboylu_2013}, it matters if the rotation is parallel or anti-parallel to the shift during tunneling which disturbs the ionization probability. Mathematically the bound state wave function has the form of $\psi^{(2p)}_{1/2\pm}(\mathbf{p},t_s)\sim 1\mp ip_{k,s}/p_{E,s}$ at the saddle point, i.e., at the moment when ionization starts. With $p_{k\,s}=-2I_p/3c$ and $p_E=-i\sqrt{2I_p}$ it follows that
\begin{eqnarray}
  \left|\psi^{(2p)}_{1/2\pm}(\mathbf{p},t_s)\right|^2\sim 1\mp \frac{2 }{3}\sqrt{\frac{2I_p}{c^2}}.
  \label{asym_psi}
\end{eqnarray}
Therefore, the ionization is preferable from the state where in the bound state the electron  rotation is opposite to the rotation of the electron due to the Lorentz force [A similar effect exists in the nonadiabatic tunneling in a circularly polarized laser field \cite{Barth_2011}, when the ionization is preferable from the state where in the bound state the electron  rotation is opposite to the rotation of the field].
Adding these two effects: the different angular momentum-magnetic field coupling in the bound state and during tunneling and momentum selective tunneling from a $p$-state,  the calculated asymmetries of Eq. (\ref{A}) are explained.

The asymmetries for $2p_{3/2}$ states have the same origin. The derived respective ionization rates are (see Fig.~\ref{B}):
\begin{eqnarray}
  \frac{dw^{(2p)}_{3/2\pm\pm}/d^3\mathbf{p}}{dw^{(2s)}/d^3\mathbf{p}}=\frac{w^{(2p)}_{3/2\pm\pm}}{w^{(2s)}}=\frac{1}{2}  \pm \frac{ \sqrt{2I_p/c^2}}{6},\label{2P3/2}\\
     \frac{dw^{(2p)}_{3/2\pm}/d^3\mathbf{p}}{dw^{(2s)}/d^3\mathbf{p}}=\frac{w^{(2p)}_{3/2\pm}}{w^{(2s)}}=\frac{1}{6} \pm \frac{\sqrt{2I_p/c^2}}{6},
     \end{eqnarray}
($I_p/c^2$ terms are neglected) which yields to asymmetries 
\begin{eqnarray}
  {\cal A}^{(2p)}_{3/2,3/2} &=& \left|\frac{dw^{(2p)}_{3/2++}/d^3\mathbf{p}-dw^{(2p)}_{3/2--}/d^3\mathbf{p}}{dw^{(2p)}_{3/2++}/d^3\mathbf{p}+dw^{(2p)}_{3/2--}/d^3\mathbf{p} } \right|\approx \frac{1}{3}\sqrt{\frac{2I_p}{c^2}}, \label{A3/2++} \\
  {\cal A}^{(2p)}_{3/2,1/2} &=& \left|\frac{dw^{(2p)}_{3/2+}/d^3\mathbf{p}-dw^{(2p)}_{3/2-}/d^3\mathbf{p}}{dw^{(2p)}_{3/2+}/d^3\mathbf{p}+dw^{(2p)}_{3/2-}/d^3\mathbf{p} }\right| \approx \sqrt{\frac{2I_p}{c^2}},
  \label{A3/2+} 
\end{eqnarray}

Here again only parts of the bound state wavefunction are allowed to tunnel that have quantum number $m_l=\pm 1$. The $2p_{3/2++}$ state is represented via the state with $m_l=1,m_s=1/2$, while $2p_{3/2--}$ state via $m_l=-1,m_s=-1/2$ (spins are opposite). Then, according to Eq. (\ref{Gamma1})
\begin{eqnarray}
 \Gamma^{(2p)}_{3/2\pm\pm} \sim e^{ -\frac{2 }{3}\frac{E_a }{E_0}}\left(1\pm  \sqrt{\frac{2I_p}{c^2}}\right), \end{eqnarray}
when $J=\pm 3/2$ and $S=\pm 1/2$, while the asymmetry due to the electron rotation in the bound state is the same as in the $J=1/2$ case, 
\begin{eqnarray}
\left|\psi^{(2p)}_{3/2\pm\pm}(\mathbf{p},t_s)\right|^2\sim 1\mp \frac{2 }{3}\sqrt{\frac{2I_p}{c^2}},
\label{asym_psi_3/2}
\end{eqnarray}
leading finally to the Eq. (\ref{A3/2++}). Similarly, from the $2p_{3/2+}$ state in the ionization contributes $m_l=1, m_s=-1/2$ state and from the $2p_{3/2-}$ does $m_l=-1, m_s=1/2$ and 
\begin{eqnarray}
 \Gamma^{(2p)}_{3/2\pm} \sim e^{ -\frac{2 }{3}\frac{E_a }{E_0}}\left(1\pm  \frac{5}{3}\sqrt{\frac{2I_p}{c^2}}\right), \end{eqnarray}
 which again leads to Eq.~(\ref{A3/2+}) taking into account Eq. (\ref{asym_psi_3/2}).

Thus, the asymmetry of ionization from a $p$-state, which is expressed by the splitting of the curves in Fig. \ref{B}, is due to the difference of the angular momentum coupling with the laser magnetic field in the bound  state and during tunneling as well as due to the fact that the ionization is larger from that bound state where the electron rotation is opposite to the electron relativistic shift along the propagation direction. However, the first effect is dominating.

The values of the ionization probability at $Z\rightarrow 0$ in Fig. \ref{B} can be easily deduced, taking into account the nonrelativistic relation between ionization probabilities of $s$- and $p$-states [mostly $m_l=\pm 1 $ states contribute to ionization] as well as the fact that the relative weight of the $m_l=1$ state in the states $2p_{3/2,3/2}$, $2p_{1/2,1/2}$, $2p_{3/2,1/2}$ are $3/2:1:1/2$, which follows from the expression of the corresponding wave functions from Eqs. (\ref{wavefunction1})-(\ref{wavefunction2}).

\section{Quantization axis  along the laser propagation direction}

\begin{figure}
  \begin{center}
    \includegraphics[width=0.4\textwidth]{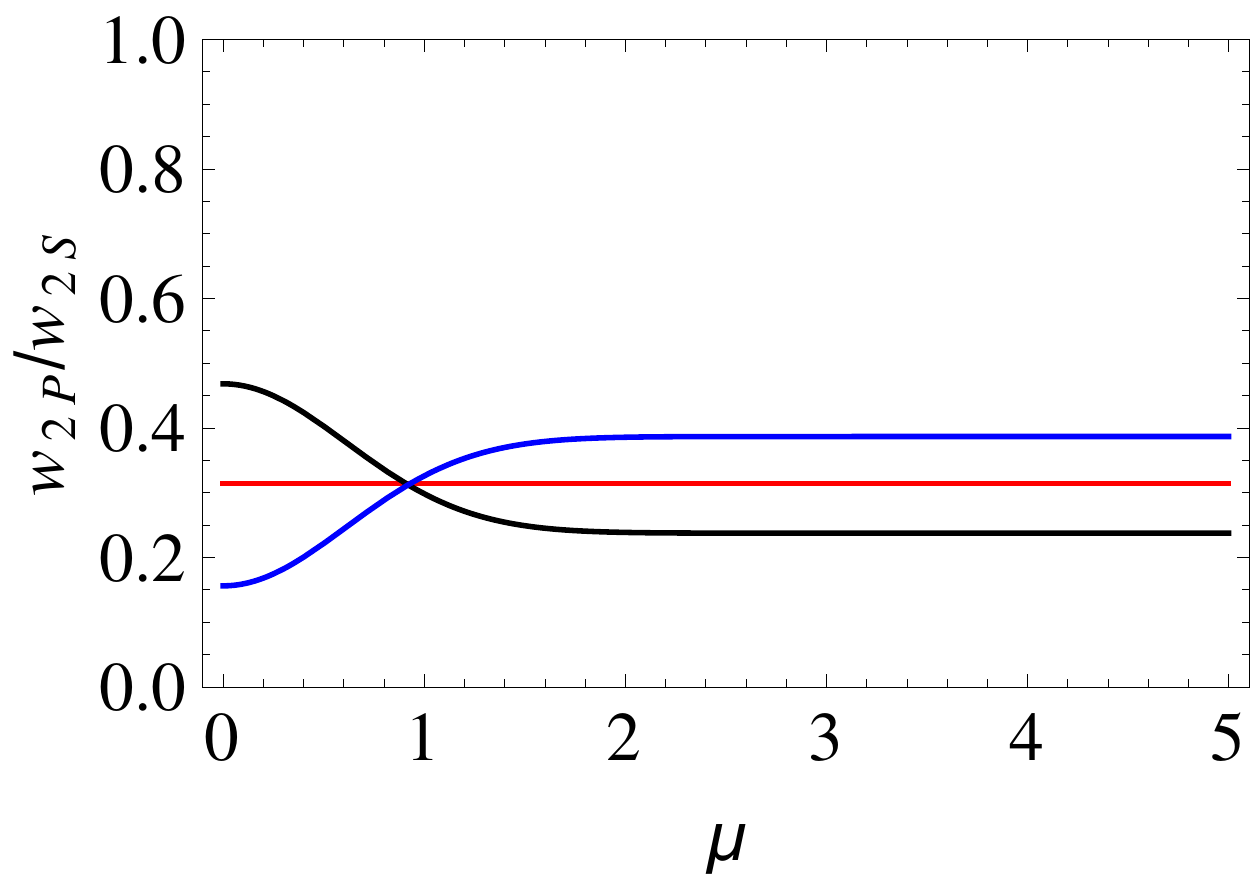}
    \caption{The relative ionization rates of $2p$-states $dw^{(2p)}/d^3\mathbf{p}$ with respect to the $2S$-state $dw^{(2s)}/d^3\mathbf{p}$ vs the laser intensity parameter $\mu=\sqrt{E_0/E_a}E_0/(c\omega)$, when the angular momentum quantization axis is along the laser propagation.  Black (top line at $\mu \rightarrow 0$): $P_{3/2,++}\rightarrow\uparrow/\downarrow$; Blue (bottom line at $\mu \rightarrow 0$): $P_{3/2,+}\rightarrow\uparrow/\downarrow$; Red (middle line at $\mu \rightarrow 0$): $P_{1/2,+}\rightarrow\uparrow/\downarrow$. The summation over the  electron final spin is indicated by $\uparrow/\downarrow$, the subscript $``++"$ indicate $m_j= 3/2$ and $``+"$ indicate $m_j= 1/2$.}
    \label{k1}
  \end{center}
\end{figure}

Now let us consider the choice of the quantization axis along the laser propagation direction. In this case
the angular momentum and the spin of the active electron is not constant before the tunneling starts, in contrast to the previous case of the quantization axis along the magnetic field and the process is altered. In particular, the spin flip becomes possible. 
Further, non-relativistically again only $m_l=\pm 1$ components of the bound state wave function are allowed to tunnel. The results for the total ionization probability of the $2p_{1/2}$-states are (see Fig. \ref{k1}):
\begin{eqnarray}
  \frac{w^{(2p)}_{1/2+}}{w^{(2s)}}\approx\frac{w^{(2p)}_{1/2-}}{w^{(2s)}}\approx \frac{1}{3} ,  
 \end{eqnarray}
where the relatively unimportant $I_p/c^2$-terms are dropped.
The ionization probabilities are almost constant which is easy to understand as follows.
The states with angular momentum up or down evolve in the bound state and mix. They can be represented by  another basis where the angular momentum is aligned along the laser magnetic field and are a linear superposition of these states. Since the ionization probabilities of these new basis states are in the leading order the same as shown in the previous section, see Eqs. (\ref{2P1/2+}), also every superposition has the same ionization probability which explains our observation.

For the $2p_{3/2}$-states the relative ionization probabilities are (see Fig. \ref{k1})
\begin{eqnarray}
  \frac{w^{(2p)}_{3/2\pm\pm}}{w^{(2s)}}=\frac{1}{4}+\frac{1}{4}\exp\left(-\frac{4 \mu^2}{3}\right)\\
  \frac{w^{(2p)}_{3/2\pm}}{w^{(2s)}}= \frac{5}{12} -\frac{1}{4} \exp\left(-\frac{4 \mu^2}{3}\right) 
 \end{eqnarray}
where $\mu=\sqrt{E_0/E_a}E_0/(c\omega)$ is the strong field parameter for the spin flip effects, see Ref. \cite{Klaiber_2014}. Here also the relatively unimportant $I_p/c^2$-dependence was dropped

One can see that the ionization probabilities for the $2p_{3/2}$-states change significantly when entering the strong field regime when $\mu\gtrsim 1$, see Fig. \ref{k1}. Note that $\mu \sim 1$ can be achieved using highly charged ions with a charge $Z\sim 20$ in a laser field with intensity $10^{21}$ W/cm$^2$. For the explanation  we express the evolving states  via the basis where the angular momentum is aligned along the magnetic field. Now the states of this basis have different ionization probabilities, see Eq. (\ref{2P3/2}),  and the total ionization probability depends on the particular superposition of these states. Moreover, the superposition depends on the bound dynamics which is different for different $\mu$, i.e., for different laser field strengths, see Eq. (\ref{dressed2P3/2}).

For instance, the state $2p_{3/2++}$ can be expressed in the basis with the quantization axis along magnetic field as follows :
\begin{eqnarray}
   |3/2++\rangle_k &=& \alpha_1|3/2++\rangle_B+\alpha_2|3/2+\rangle_B\nonumber\\
&+&\alpha_3|3/2-\rangle_B+\alpha_4|3/2--\rangle_B,
\end{eqnarray}
with $\alpha_1=1/(2\sqrt{2})\exp[iA/c]$, $\alpha_2=\sqrt{3}/(2\sqrt{2})\exp[iA/6c]$, $\alpha_3=\sqrt{3}/(2\sqrt{2})\exp[-iA/6c]$ and $\alpha_4=1/(2\sqrt{2})\exp[-iA/c]$.
The states $|3/2++\rangle_B$ and $|3/2+\rangle_B$ have different ionization probabilities,  the ratio  of the probabilities is $W^{(2p)}_{3/2++;B}:W^{(2p)}_{3/2+ ;B}=1/2:1/6$, see Eq. (\ref{2P3/2}), and the field dependent phases due to Zeeman splitting are different as well. Therefore, the ionization probability is essentially field dependent.

Whereas for small fields strength the phases are negligible $A/c\ll 1$ ($\mu\ll 1$), for large field strength they average out. Because of that, at $\mu\gg 1$ the total ionization rate of the  $|3/2++\rangle_k$ state can  be given by:
\begin{eqnarray}
  W^{(2p)}_{3/2++}\approx \sum_{i=1}^4|\alpha_i|^2|M_i|^2=\frac{1}{4}, 
 \end{eqnarray}
with $|M_1|^2\approx |M_4|^2=W^{(2p)}_{3/2++;B}=1/2$ and $|M_2|^2\approx |M_3|^2=W^{(2p)}_{3/2+;B}=1/6$, explaining the $\mu\gg 1$ asymptotic behaviour of the black line in Fig.~\ref{k1}. Similarly can be explained the $\mu\gg 1$ asymptotics of the other curves.

In the weak field limit $\mu \ll 1$ the phases $\sim A/c$ determining the angular momentum precession in the magnetic field can be neglected. In this case the total ionization rate for the $|3/2++\rangle_k$ state can  be given by:
\begin{eqnarray}
  W^{(2p)}_{3/2++}&\approx& |\alpha_1 M_1 +\alpha_3 M_3|^2+|\alpha_2 M_2 +\alpha_4M_4|^2\\
  &&\approx \left|\frac{1}{4}+\frac{1}{4}\right|^2+\left|\frac{1}{4}+\frac{1}{4}\right|^2=\frac{1}{2}.
  \label{large_mu}
\end{eqnarray}
which explains the $\mu \ll 1$ asymptotics of the black line in Fig.~\ref{k1}. Here we note that the states $|Jm_j\rangle_B$ do not precess in the laser field and ionize into states with spin up with respect to the magnetic field, when $m_j=3/2$ or $m_j=-1/2$ or spin down in the case $m_j=1/2$ or $m_j=-3/2$, respectively. That is why in Eq. (\ref{large_mu}) $M_1$ interferes only with $M_3$, and $M_2$ interferes only with $M_4$.  

Thus, in weak laser fields there is interference in the ionization probability from the  superposition of states, while in strong fields the interference is wiped out because of the fast precession of the bound state. This has a consequence that $m_j$ dependence of the ioniztion rate is different in weak and strong field asymptotics, see Fig.~\ref{k1}.

\section{ Quantization axis along the laser electric field}

\begin{figure}
  \begin{center}
    \includegraphics[width=0.4\textwidth]{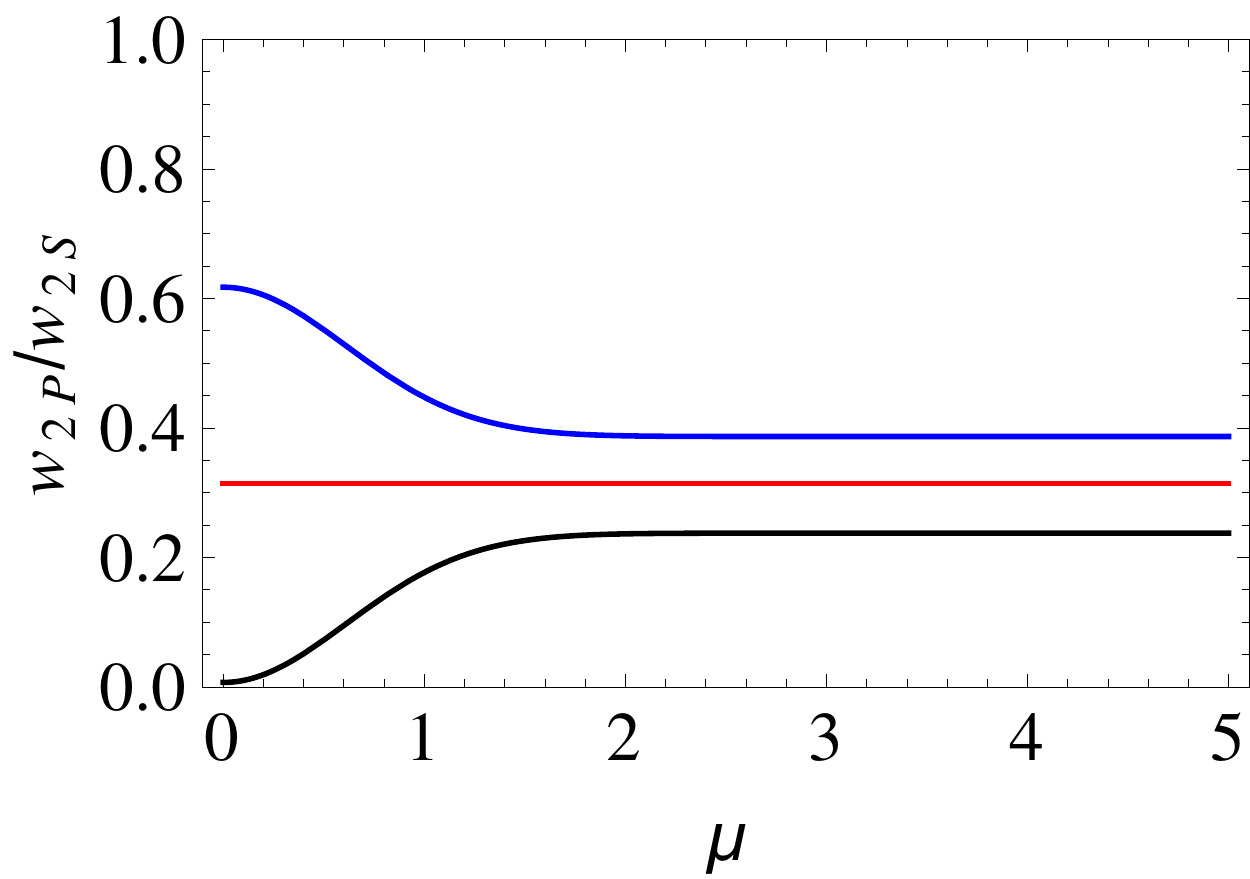}
    \caption{The relative ionization rates of $2p$-states $dw^{(2p)}/d^3\mathbf{p}$ with respect to the $2S$-state $dw^{(2s)}/d^3\mathbf{p}$ vs the laser intensity parameter $\mu=\sqrt{E_0/E_a}E_0/(c\omega)$, when the angular momentum quantization axis is along the laser electric field direction.  Black (bottom line): $P_{3/2++}\rightarrow\uparrow/\downarrow$; Blue (top line): $P_{3/2+}\rightarrow\uparrow/\downarrow$; Red (middle line): $P_{1/2,+}\rightarrow\uparrow/\downarrow$. The summation over the  electron final spin is indicated by $\uparrow/\downarrow$, the subscript $``++"$ indicate $m_j= 3/2$ and $``+"$ indicate $m_j= 1/2$.}
    \label{E1}
  \end{center}
\end{figure}
Finally, let us consider the case when the quantization axis is along the laser electric field.
Here non-relativistically parts of the bound state wave function with quantum number $m_l=0$ are allowed to tunnel through the barrier. Therefore, the $2p_{3/2++}$ and $2p_{3/2--}$ states have zero ionization probability for weak laser fields, see Fig. \ref{E1}.

For the $2p_{3/2}$-states the relative ionization probabilities are (see Fig.~\ref{E1})
\begin{eqnarray}
  \frac{w^{(2p)}_{3/2\pm\pm}}{w^{(2s)}}\approx \frac{1}{4}-\frac{1}{4}\exp\left(-\frac{4 \mu^2}{3}\right)\\
  \frac{w^{(2p)}_{3/2\pm}}{w^{(2s)}}\approx \frac{5}{12} +\frac{1}{4}\exp\left(-\frac{4 \mu^2}{3}\right),
  \end{eqnarray}
where again $I_p/c^2$-corrections are dropped. The intuitive explanation of the asymptotic behaviour of the ionization probabilities for small and large laser field strength can be done analogous to that of the previous section.

\section{Conclusion}

We have investigated the tunneling ionization of an highly charged ion from an excited $p$-state of an hydrogenlike ion in a linearly polarized strong laser field in the relativistic regime. The ionization picture is analysed in three possible setups for the angular momentum and spin quantization axis. When the quantization axis is along the laser magnetic field, then there is no spin flip effect but there exists a large spin asymmetry. This is in contrast to the case of the ionization from an $s$-state where the spin asymmetry is vanishing. The spin asymmetry of the $p$-state ionization is due to two reasons. Firstly, there is a difference in the Zeeman splitting of the electron energy  in the bound state and during tunneling. Secondly, in the relativistic regime the tunneling electron acquires a shift along the laser propagation direction during the under-the barrier motion due to the $\textbf{v} \times \textbf{B}$ force. The ionization probability is larger for those states (for such magnetic quantum number) where the electron rotation in the bound state in the $(\textbf{k},\textbf{E})$-plane is opposite to the  relativistic shift.  Because at a certain value of the projection of the total angular momentum the spin states are entangled with the states of the magnetic quantum number, the mentioned asymmetry with respect to the magnetic quantum number is observed as a spin asymmetry.

In the case when the spin quantization axis is along the laser propagation direction (or along the laser electric field), we have investigated the dependence of the ionization rate on the laser intensity  for different projections of the total angular momentum $m_j$. The dependence of the ionization rate on the projection of the total angular momentum appears to be different for weak and strong fields (nonrelativistic and relativistic regimes). Moreover, the $m_j$-dependence of the ionization rate in the relativistic regime is reverted with respect to the case of the nonrelativistic regime. We have identified the intensity parameter $\mu=\sqrt{E_0/E_a}(E_0/c\omega)$ which governs this behaviour. Correspondingly, the intensity dependence of the ionization rate are different for the states with different projections of the total angular momentum. This effect can be observed using highly charged ions with a charge $Z\sim 20$ in a laser field with intensity $10^{21}$ W/cm$^2$. We have provided the intuitive description of these properties. The state with a certain total angular momentum along the laser propagation (field) direction can be represented as a superposition of states  with different projections of the total angular momentum on the magnetic field direction. In weak laser fields there is interference in the ionization probability from the above mentioned superposition of states. Meanwhile, in strong fields the fast precession of the bound state destroys the interference in ionization.

\section*{Acknowledgement} 
We gratefully acknowledge the helpful discussions with Prof. C. H. Keitel. This research was  supported in part by the National Science Foundation under Grant No. PHY11-25915.

\bibliography{strong_fields_bibliography}

\begin{thebibliography}{10}%
\makeatletter
\providecommand \@ifxundefined [1]{%
 \ifx #1\undefined \expandafter \@firstoftwo
 \else \expandafter \@secondoftwo
\fi
}%
\providecommand \@ifnum [1]{%
 \ifnum #1\expandafter \@firstoftwo
 \else \expandafter \@secondoftwo
\fi
}%
\providecommand \enquote [1]{``#1''}%
\providecommand \bibnamefont  [1]{#1}%
\providecommand \bibfnamefont [1]{#1}%
\providecommand \citenamefont [1]{#1}%
\providecommand\href[0]{\@sanitize\@href}%
\providecommand\@href[1]{\endgroup\@@startlink{#1}\endgroup\@@href}%
\providecommand\@@href[1]{#1\@@endlink}%
\providecommand \@sanitize [0]{\begingroup\catcode`\&12\catcode`\#12\relax}%
\@ifxundefined \pdfoutput {\@firstoftwo}{%
 \@ifnum{\z@=\pdfoutput}{\@firstoftwo}{\@secondoftwo}%
}{%
 \providecommand\@@startlink[1]{\leavevmode\special{html:<a href="#1">}}%
 \providecommand\@@endlink[0]{\special{html:</a>}}%
}{%
 \providecommand\@@startlink[1]{%
  \leavevmode
  \pdfstartlink
   attr{/Border[0 0 1 ]/H/I/C[0 1 1]}%
   user{/Subtype/Link/A<</Type/Action/S/URI/URI(#1)>>}%
  \relax
 }%
 \providecommand\@@endlink[0]{\pdfendlink}%
}%
\providecommand \url  [0]{\begingroup\@sanitize \@url }%
\providecommand \@url [1]{\endgroup\@href {#1}{\urlprefix}}%
\providecommand \urlprefix [0]{URL }%
\providecommand \Eprint[0]{\href }%
\@ifxundefined \urlstyle {%
  \providecommand \doi [1]{doi:\discretionary{}{}{}#1}%
}{%
  \providecommand \doi [0]{doi:\discretionary{}{}{}\begingroup
  \urlstyle{rm}\Url }%
}%
\providecommand \doibase [0]{http://dx.doi.org/}%
\providecommand \Doi[1]{\href{\doibase#1}}%
\providecommand \bibAnnote [3]{%
  \BibitemShut{#1}%
  \begin{quotation}\noindent
    \textsc{Key:}\ #2\\\textsc{Annotation:}\ #3%
  \end{quotation}%
}%
\providecommand \bibAnnoteFile [2]{%
  \IfFileExists{#2}{\bibAnnote {#1} {#2} {\input{#2}}}{}%
}%
\providecommand \typeout [0]{\immediate \write \m@ne }%
\providecommand \selectlanguage [0]{\@gobble}%
\providecommand \bibinfo [0]{\@secondoftwo}%
\providecommand \bibfield [0]{\@secondoftwo}%
\providecommand \translation [1]{[#1]}%
\providecommand \BibitemOpen[0]{}%
\providecommand \bibitemStop [0]{}%
\providecommand \bibitemNoStop [0]{.\EOS\space}%
\providecommand \EOS [0]{\spacefactor3000\relax}%
\providecommand \BibitemShut [1]{\csname bibitem#1\endcsname}%
\bibitem{DiChiara_2008}%
  \BibitemOpen
  \bibfield{author}{%
  \bibinfo {author} {\bibfnamefont{A.~D.}\ \bibnamefont{DiChiara}}, \bibinfo
  {author} {\bibfnamefont{I.}~\bibnamefont{Ghebregziabher}}, \bibinfo {author}
  {\bibfnamefont{R.}~\bibnamefont{Sauer}}, \bibinfo {author}
  {\bibfnamefont{J.}~\bibnamefont{Waesche}}, \bibinfo {author}
  {\bibfnamefont{S.}~\bibnamefont{Palaniyappan}}, \bibinfo {author}
  {\bibfnamefont{B.~L.}\ \bibnamefont{Wen}},\ and\ \bibinfo {author}
  {\bibfnamefont{B.~C.}\ \bibnamefont{Walker}},\ }%
  \bibfield{journal}{%
  \bibinfo {journal} {Phys. Rev. Lett.}\ }%
  \textbf{\bibinfo {volume} {101}},\ \bibinfo {pages} {173002} (\bibinfo {year}
  {2008})%
  \bibAnnoteFile{NoStop}{DiChiara_2008}%
\bibitem{Palaniyappan_2008}%
  \BibitemOpen
  \bibfield{author}{%
  \bibinfo {author} {\bibfnamefont{S.}~\bibnamefont{Palaniyappan}}, \bibinfo
  {author} {\bibfnamefont{R.}~\bibnamefont{Mitchell}}, \bibinfo {author}
  {\bibfnamefont{R.}~\bibnamefont{Sauer}}, \bibinfo {author}
  {\bibfnamefont{I.}~\bibnamefont{Ghebregziabher}}, \bibinfo {author}
  {\bibfnamefont{S.~L.}\ \bibnamefont{White}}, \bibinfo {author}
  {\bibfnamefont{M.~F.}\ \bibnamefont{Decamp}},\ and\ \bibinfo {author}
  {\bibfnamefont{B.~C.}\ \bibnamefont{Walker}},\ }%
  \bibfield{journal}{%
  \bibinfo {journal} {Phys. Rev. Lett.}\ }%
  \textbf{\bibinfo {volume} {100}},\ \bibinfo {pages} {183001} (\bibinfo {year}
  {2008})%
  \bibAnnoteFile{NoStop}{Palaniyappan_2008}%
\bibitem{DiChiara_2010}%
  \BibitemOpen
  \bibfield{author}{%
  \bibinfo {author} {\bibfnamefont{A.~D.}\ \bibnamefont{DiChiara}}, \bibinfo
  {author} {\bibfnamefont{I.}~\bibnamefont{Ghebregziabher}}, \bibinfo {author}
  {\bibfnamefont{J.~M.}\ \bibnamefont{Waesche}}, \bibinfo {author}
  {\bibfnamefont{T.}~\bibnamefont{Stanev}}, \bibinfo {author}
  {\bibfnamefont{N.}~\bibnamefont{Ekanayake}}, \bibinfo {author}
  {\bibfnamefont{L.~R.}\ \bibnamefont{Barclay}}, \bibinfo {author}
  {\bibfnamefont{S.~J.}\ \bibnamefont{Wells}}, \bibinfo {author}
  {\bibfnamefont{A.}~\bibnamefont{Watts}}, \bibinfo {author}
  {\bibfnamefont{M.}~\bibnamefont{Videtto}}, \bibinfo {author}
  {\bibfnamefont{C.~A.}\ \bibnamefont{Mancuso}},\ and\ \bibinfo {author}
  {\bibfnamefont{B.~C.}\ \bibnamefont{Walker}},\ }%
  \bibfield{journal}{%
  \bibinfo {journal} {Phys. Rev. A}\ }%
  \textbf{\bibinfo {volume} {81}},\ \bibinfo {pages} {043417} (\bibinfo {year}
  {2010})%
  \bibAnnoteFile{NoStop}{DiChiara_2010}%
\bibitem{Ekanayake_2013}%
  \BibitemOpen
  \bibfield{author}{%
  \bibinfo {author} {\bibfnamefont{N.}~\bibnamefont{Ekanayake}}, \bibinfo
  {author} {\bibfnamefont{S.}~\bibnamefont{Luo}}, \bibinfo {author}
  {\bibfnamefont{P.~D.}\ \bibnamefont{Grugan}}, \bibinfo {author}
  {\bibfnamefont{W.~B.}\ \bibnamefont{Crosby}}, \bibinfo {author}
  {\bibfnamefont{A.~D.}\ \bibnamefont{Camilo}}, \bibinfo {author}
  {\bibfnamefont{C.~V.}\ \bibnamefont{McCowan}}, \bibinfo {author}
  {\bibfnamefont{R.}~\bibnamefont{Scalzi}}, \bibinfo {author}
  {\bibfnamefont{A.}~\bibnamefont{Tramontozzi}}, \bibinfo {author}
  {\bibfnamefont{L.~E.}\ \bibnamefont{Howard}}, \bibinfo {author}
  {\bibfnamefont{S.~J.}\ \bibnamefont{Wells}}, \bibinfo {author}
  {\bibfnamefont{C.}~\bibnamefont{Mancuso}}, \bibinfo {author}
  {\bibfnamefont{T.}~\bibnamefont{Stanev}}, \bibinfo {author}
  {\bibfnamefont{M.~F.}\ \bibnamefont{Decamp}},\ and\ \bibinfo {author}
  {\bibfnamefont{B.~C.}\ \bibnamefont{Walker}},\ }%
  \bibfield{journal}{%
  \bibinfo {journal} {Phys. Rev. Lett.}\ }%
  \textbf{\bibinfo {volume} {110}},\ \bibinfo {pages} {203003} (\bibinfo {year}
  {2013})%
  \bibAnnoteFile{NoStop}{Ekanayake_2013}%
\bibitem{RMP_2012}%
  \BibitemOpen
  \bibfield{author}{%
  \bibinfo {author} {\bibfnamefont{A.}~\bibnamefont{{Di Piazza}}}, \bibinfo
  {author} {\bibfnamefont{C.}~\bibnamefont{M\"uller}}, \bibinfo {author}
  {\bibfnamefont{K.~Z.}\ \bibnamefont{Hatsagortsyan}},\ and\ \bibinfo {author}
  {\bibfnamefont{C.~H.}\ \bibnamefont{Keitel}},\ }%
  \bibfield{journal}{%
  \bibinfo {journal} {Rev. Mod. Phys.}\ }%
  \textbf{\bibinfo {volume} {84}},\ \bibinfo {pages} {1177} (\bibinfo {year}
  {2012})%
  \bibAnnoteFile{NoStop}{RMP_2012}%
\bibitem{Klaiber_2013c}%
  \BibitemOpen
  \bibfield{author}{%
  \bibinfo {author} {\bibfnamefont{M.}~\bibnamefont{Klaiber}}, \bibinfo
  {author} {\bibfnamefont{E.}~\bibnamefont{Yakaboylu}}, \bibinfo {author}
  {\bibfnamefont{H.}~\bibnamefont{Bauke}}, \bibinfo {author}
  {\bibfnamefont{K.~Z.}\ \bibnamefont{Hatsagortsyan}},\ and\ \bibinfo {author}
  {\bibfnamefont{C.~H.}\ \bibnamefont{Keitel}},\ }%
  \bibfield{journal}{%
  \bibinfo {journal} {Phys. Rev. Lett.}\ }%
  \textbf{\bibinfo {volume} {110}},\ \bibinfo {pages} {153004} (\bibinfo {year}
  {2013})%
  \bibAnnoteFile{NoStop}{Klaiber_2013c}%
\bibitem{Yakaboylu_2013}%
  \BibitemOpen
  \bibfield{author}{%
  \bibinfo {author} {\bibfnamefont{E.}~\bibnamefont{Yakaboylu}}, \bibinfo
  {author} {\bibfnamefont{M.}~\bibnamefont{Klaiber}}, \bibinfo {author}
  {\bibfnamefont{H.}~\bibnamefont{Bauke}}, \bibinfo {author}
  {\bibfnamefont{K.~Z.}\ \bibnamefont{Hatsagortsyan}},\ and\ \bibinfo {author}
  {\bibfnamefont{C.~H.}\ \bibnamefont{Keitel}},\ }%
  \bibfield{journal}{%
  \bibinfo {journal} {Phys. Rev. A}\ }%
  \textbf{\bibinfo {volume} {88}},\ \bibinfo {pages} {063421} (\bibinfo {year}
  {2013})%
  \bibAnnoteFile{NoStop}{Yakaboylu_2013}%
\bibitem{Yakaboylu_2014b}%
  \BibitemOpen
  \bibfield{author}{%
  \bibinfo {author} {\bibfnamefont{E.}~\bibnamefont{Yakaboylu}}, \bibinfo
  {author} {\bibfnamefont{M.}~\bibnamefont{Klaiber}},\ and\ \bibinfo {author}
  {\bibfnamefont{K.~Z.}\ \bibnamefont{Hatsagortsyan}},\ }%
  \bibfield{journal}{%
  \bibinfo {journal} {Phys. Rev. A}\ }%
  \textbf{\bibinfo {volume} {90}},\ \bibinfo {pages} {012116} (\bibinfo {year}
  {2014})%
  \bibAnnoteFile{NoStop}{Yakaboylu_2014b}%
\bibitem{Ahrens_2012}%
  \BibitemOpen
  \bibfield{author}{%
  \bibinfo {author} {\bibfnamefont{S.}~\bibnamefont{Ahrens}}, \bibinfo {author}
  {\bibfnamefont{H.}~\bibnamefont{Bauke}}, \bibinfo {author}
  {\bibfnamefont{C.~H.}\ \bibnamefont{Keitel}},\ and\ \bibinfo {author}
  {\bibfnamefont{C.}~\bibnamefont{M\"uller}},\ }%
  \bibfield{journal}{%
  \bibinfo {journal} {Phys. Rev. Lett.}\ }%
  \textbf{\bibinfo {volume} {109}},\ \bibinfo {pages} {043601} (\bibinfo {year}
  {2012})%
  \bibAnnoteFile{NoStop}{Ahrens_2012}%
\bibitem{Ahrens_2013}%
  \BibitemOpen
  \bibfield{author}{%
  \bibinfo {author} {\bibfnamefont{S.}~\bibnamefont{Ahrens}}, \bibinfo {author}
  {\bibfnamefont{H.}~\bibnamefont{Bauke}}, \bibinfo {author}
  {\bibfnamefont{C.~H.}\ \bibnamefont{Keitel}},\ and\ \bibinfo {author}
  {\bibfnamefont{C.}~\bibnamefont{M\"uller}},\ }%
  \bibfield{journal}{%
  \bibinfo {journal} {Phys. Rev. A}\ }%
  \textbf{\bibinfo {volume} {88}},\ \bibinfo {pages} {012115} (\bibinfo {year}
  {2013})%
  \bibAnnoteFile{NoStop}{Ahrens_2013}%
\bibitem{Skoromnik_2013}%
  \BibitemOpen
  \bibfield{author}{%
  \bibinfo {author} {\bibfnamefont{O.~D.}\ \bibnamefont{Skoromnik}}, \bibinfo
  {author} {\bibfnamefont{I.~D.}\ \bibnamefont{Feranchuk}},\ and\ \bibinfo
  {author} {\bibfnamefont{C.~H.}\ \bibnamefont{Keitel}},\ }%
  \bibfield{journal}{%
  \bibinfo {journal} {Phys. Rev. A}\ }%
  \textbf{\bibinfo {volume} {87}},\ \bibinfo {pages} {052107} (\bibinfo {year}
  {2013})%
  \bibAnnoteFile{NoStop}{Skoromnik_2013}%
\bibitem{Wen_2014}%
  \BibitemOpen
  \bibfield{author}{%
  \bibinfo {author} {\bibfnamefont{C.~H.~K.}\ \bibnamefont{Meng~Wen},
  \bibfnamefont{Heiko~Bauke}},\ }%
  \bibfield{journal}{%
  \bibinfo {journal} {arXiv},\ \bibinfo {pages} {1406.3659}}%
   (\bibinfo {year} {2014}),\ \bibinfo {note} {{[physics.plasm-ph]}}%
  \bibAnnoteFile{NoStop}{Wen_2014}%
\bibitem{Ahrens_2014}%
  \BibitemOpen
  \bibfield{author}{%
  \bibinfo {author} {\bibfnamefont{S.}~\bibnamefont{Ahrens}}, \bibinfo {author}
  {\bibfnamefont{H.}~\bibnamefont{Bauke}}, \bibinfo {author}
  {\bibfnamefont{C.~H.}\ \bibnamefont{Keitel}},\ and\ \bibinfo {author}
  {\bibfnamefont{R.}~\bibnamefont{Grobe}},\ }%
  \bibfield{journal}{%
  \bibinfo {journal} {arXiv},\ \bibinfo {pages} {1401.5976}}%
   (\bibinfo {year} {2014}),\ \bibinfo {note} {{[quant-ph]}}%
  \bibAnnoteFile{NoStop}{Ahrens_2014}%
\bibitem{Bauke_2014a}%
  \BibitemOpen
  \bibfield{author}{%
  \bibinfo {author} {\bibfnamefont{H.}~\bibnamefont{Bauke}}, \bibinfo {author}
  {\bibfnamefont{S.}~\bibnamefont{Ahrens}}, \bibinfo {author}
  {\bibfnamefont{C.~H.}\ \bibnamefont{Keitel}},\ and\ \bibinfo {author}
  {\bibfnamefont{R.}~\bibnamefont{Grobe}},\ }%
  \bibfield{journal}{%
  \bibinfo {journal} {Phys. Rev. A}\ }%
  \textbf{\bibinfo {volume} {89}},\ \bibinfo {pages} {052101} (\bibinfo {year}
  {2014})%
  \bibAnnoteFile{NoStop}{Bauke_2014a}%
\bibitem{Bauke_2014b}%
  \BibitemOpen
  \bibfield{author}{%
  \bibinfo {author} {\bibfnamefont{H.}~\bibnamefont{Bauke}}, \bibinfo {author}
  {\bibfnamefont{S.}~\bibnamefont{Ahrens}}, \bibinfo {author}
  {\bibfnamefont{C.~H.}\ \bibnamefont{Keitel}},\ and\ \bibinfo {author}
  {\bibfnamefont{R.}~\bibnamefont{Grobe}},\ }%
  \bibfield{journal}{%
  \bibinfo {journal} {New J. Phys.}\ }%
  \textbf{\bibinfo {volume} {16}},\ \bibinfo {pages} {043012} (\bibinfo {year}
  {2014})%
  \bibAnnoteFile{NoStop}{Bauke_2014b}%
\bibitem{Walser_2002}%
  \BibitemOpen
  \bibfield{author}{%
  \bibinfo {author} {\bibfnamefont{M.~W.}\ \bibnamefont{Walser}}, \bibinfo
  {author} {\bibfnamefont{D.~J.}\ \bibnamefont{Urbach}}, \bibinfo {author}
  {\bibfnamefont{K.~Z.}\ \bibnamefont{Hatsagortsyan}}, \bibinfo {author}
  {\bibfnamefont{S.~X.}\ \bibnamefont{Hu}},\ and\ \bibinfo {author}
  {\bibfnamefont{C.~H.}\ \bibnamefont{Keitel}},\ }%
  \bibfield{journal}{%
  \bibinfo {journal} {Phys. Rev. A}\ }%
  \textbf{\bibinfo {volume} {65}},\ \bibinfo {pages} {043410} (\bibinfo {year}
  {2002})%
  \bibAnnoteFile{NoStop}{Walser_2002}%
\bibitem{Hu_1999}%
  \BibitemOpen
  \bibfield{author}{%
  \bibinfo {author} {\bibfnamefont{S.~X.}\ \bibnamefont{Hu}}\ and\ \bibinfo
  {author} {\bibfnamefont{C.~H.}\ \bibnamefont{Keitel}},\ }%
  \bibfield{journal}{%
  \bibinfo {journal} {Phys. Rev. Lett.}\ }%
  \textbf{\bibinfo {volume} {83}},\ \bibinfo {pages} {4709} (\bibinfo {year}
  {1999})%
  \bibAnnoteFile{NoStop}{Hu_1999}%
\bibitem{Walser_2001}%
  \BibitemOpen
  \bibfield{author}{%
  \bibinfo {author} {\bibfnamefont{M.~W.}\ \bibnamefont{Walser}}\ and\ \bibinfo
  {author} {\bibfnamefont{C.~H.}\ \bibnamefont{Keitel}},\ }%
  \bibfield{journal}{%
  \bibinfo {journal} {Opt. Commun.}\ }%
  \textbf{\bibinfo {volume} {199}},\ \bibinfo {pages} {447 } (\bibinfo {year}
  {2001})%
  \bibAnnoteFile{NoStop}{Walser_2001}%
\bibitem{Faisal_2004}%
  \BibitemOpen
  \bibfield{author}{%
  \bibinfo {author} {\bibfnamefont{F.~H.~M.}\ \bibnamefont{Faisal}}\ and\
  \bibinfo {author} {\bibfnamefont{S.}~\bibnamefont{Bhattacharyya}},\ }%
  \bibfield{journal}{%
  \bibinfo {journal} {Phys. Rev. Lett.}\ }%
  \textbf{\bibinfo {volume} {93}},\ \bibinfo {pages} {053002} (\bibinfo {year}
  {2004})%
  \bibAnnoteFile{NoStop}{Faisal_2004}%
\bibitem{Klaiber_2014}%
  \BibitemOpen
  \bibfield{author}{%
  \bibinfo {author} {\bibfnamefont{M.}~\bibnamefont{Klaiber}}, \bibinfo
  {author} {\bibfnamefont{E.}~\bibnamefont{Yakaboylu}}, \bibinfo {author}
  {\bibfnamefont{C.}~\bibnamefont{M\"uller}}, \bibinfo {author}
  {\bibfnamefont{H.}~\bibnamefont{Bauke}}, \bibinfo {author}
  {\bibfnamefont{G.~G.}\ \bibnamefont{Paulus}},\ and\ \bibinfo {author}
  {\bibfnamefont{K.~Z.}\ \bibnamefont{Hatsagortsyan}},\ }%
  \bibfield{journal}{%
  \bibinfo {journal} {J. Phys. B}\ }%
  \textbf{\bibinfo {volume} {47}},\ \bibinfo {pages} {065603} (\bibinfo {year}
  {2014})%
  \bibAnnoteFile{NoStop}{Klaiber_2014}%
\bibitem{Bhattacharyya_2007}%
  \BibitemOpen
  \bibfield{author}{%
  \bibinfo {author} {\bibfnamefont{S.}~\bibnamefont{Bhattacharyya}}, \bibinfo
  {author} {\bibfnamefont{M.}~\bibnamefont{M.}}, \bibinfo {author}
  {\bibfnamefont{J.}~\bibnamefont{Chakrabarti}},\ and\ \bibinfo {author}
  {\bibfnamefont{F.~H.~M.}\ \bibnamefont{Faisal}},\ }%
  \bibfield{journal}{%
  \bibinfo {journal} {J. Phys. Conf. Ser.}\ }%
  \textbf{\bibinfo {volume} {80}},\ \bibinfo {pages} {012029} (\bibinfo {year}
  {2007})%
  \bibAnnoteFile{NoStop}{Bhattacharyya_2007}%
\bibitem{Bhattacharyya_2011}%
  \BibitemOpen
  \bibfield{author}{%
  \bibinfo {author} {\bibfnamefont{S.}~\bibnamefont{Bhattacharyya}}, \bibinfo
  {author} {\bibfnamefont{M.}~\bibnamefont{Mazumder}}, \bibinfo {author}
  {\bibfnamefont{J.}~\bibnamefont{Chakrabarti}},\ and\ \bibinfo {author}
  {\bibfnamefont{F.~H.~M.}\ \bibnamefont{Faisal}},\ }%
  \bibfield{journal}{%
  \bibinfo {journal} {Phys. Rev. A}\ }%
  \textbf{\bibinfo {volume} {83}},\ \bibinfo {pages} {043407} (\bibinfo {year}
  {2011})%
  \bibAnnoteFile{NoStop}{Bhattacharyya_2011}%
\bibitem{Perelomov_1966a}%
  \BibitemOpen
  \bibfield{author}{%
  \bibinfo {author} {\bibfnamefont{A.~M.}\ \bibnamefont{Perelomov}}\ and\
  \bibinfo {author} {\bibfnamefont{V.~S.}\ \bibnamefont{Popov}},\ }%
  \bibfield{journal}{%
  \bibinfo {journal} {Zh. Exp. Theor. Fiz.}\ }%
  \textbf{\bibinfo {volume} {50}},\ \bibinfo {pages} {1393} (\bibinfo {year}
  {1966})%
  \bibAnnoteFile{NoStop}{Perelomov_1966a}%
\bibitem{Perelomov_1966b}%
  \BibitemOpen
  \bibfield{author}{%
  \bibinfo {author} {\bibfnamefont{A.~M.}\ \bibnamefont{Perelomov}}, \bibinfo
  {author} {\bibfnamefont{V.~S.}\ \bibnamefont{Popov}},\ and\ \bibinfo {author}
  {\bibfnamefont{V.~M.}\ \bibnamefont{Terent'ev}},\ }%
  \bibfield{journal}{%
  \bibinfo {journal} {Zh. Exp. Theor. Fiz.}\ }%
  \textbf{\bibinfo {volume} {51}},\ \bibinfo {pages} {309} (\bibinfo {year}
  {1966})%
  \bibAnnoteFile{NoStop}{Perelomov_1966b}%
\bibitem{Perelomov_1967a}%
  \BibitemOpen
  \bibfield{author}{%
  \bibinfo {author} {\bibfnamefont{A.~M.}\ \bibnamefont{Perelomov}}\ and\
  \bibinfo {author} {\bibfnamefont{V.~S.}\ \bibnamefont{Popov}},\ }%
  \bibfield{journal}{%
  \bibinfo {journal} {Zh. Exp. Theor. Fiz.}\ }%
  \textbf{\bibinfo {volume} {52}},\ \bibinfo {pages} {514} (\bibinfo {year}
  {1967})%
  \bibAnnoteFile{NoStop}{Perelomov_1967a}%
\bibitem{Ammosov_1986}%
  \BibitemOpen
  \bibfield{author}{%
  \bibinfo {author} {\bibfnamefont{M.~V.}\ \bibnamefont{Ammosov}}, \bibinfo
  {author} {\bibfnamefont{N.~B.}\ \bibnamefont{Delone}},\ and\ \bibinfo
  {author} {\bibfnamefont{V.~P.}\ \bibnamefont{Krainov}},\ }%
  \bibfield{journal}{%
  \bibinfo {journal} {Zh. Eksp. Teor. Fiz.}\ }%
  \textbf{\bibinfo {volume} {91}},\ \bibinfo {pages} {2008} (\bibinfo {year}
  {1986})%
  \bibAnnoteFile{NoStop}{Ammosov_1986}%
\bibitem{Popov_2004}%
  \BibitemOpen
  \bibfield{author}{%
  \bibinfo {author} {\bibfnamefont{V.~S.}\ \bibnamefont{Popov}}, \bibinfo
  {author} {\bibfnamefont{B.~M.}\ \bibnamefont{Karnakov}},\ and\ \bibinfo
  {author} {\bibfnamefont{V.~D.}\ \bibnamefont{Mur}},\ }%
  \bibfield{journal}{%
  \bibinfo {journal} {Zh. Exp. Theor. Fiz.}\ }%
  \textbf{\bibinfo {volume} {79}},\ \bibinfo {pages} {320} (\bibinfo {year}
  {2004})%
  \bibAnnoteFile{NoStop}{Popov_2004}%
\bibitem{Zon_2005}%
  \BibitemOpen
  \bibfield{author}{%
  \bibinfo {author} {\bibfnamefont{B.~A.}\ \bibnamefont{Zon}}\ and\ \bibinfo
  {author} {\bibfnamefont{A.~S.}\ \bibnamefont{Kornev}},\ }%
  \bibfield{journal}{%
  \bibinfo {journal} {J. Exp. Theor. Phys.}\ }%
  \textbf{\bibinfo {volume} {101}},\ \bibinfo {pages} {1009} (\bibinfo {year}
  {2005})%
  \bibAnnoteFile{NoStop}{Zon_2005}%
\bibitem{Liu_2012}%
  \BibitemOpen
  \bibfield{author}{%
  \bibinfo {author} {\bibfnamefont{C.}~\bibnamefont{Liu}}\ and\ \bibinfo
  {author} {\bibfnamefont{M.}~\bibnamefont{Nisoli}},\ }%
  \bibfield{journal}{%
  \Doi{10.1103/PhysRevA.85.013418}{\bibinfo {journal} {Phys. Rev. A}}\ }%
  \textbf{\bibinfo {volume} {85}},\ \bibinfo {pages} {013418} (\bibinfo {month}
  {Jan}\ \bibinfo {year} {2012}),\
  \url{http://link.aps.org/doi/10.1103/PhysRevA.85.013418}%
  \bibAnnoteFile{NoStop}{Liu_2012}%
\bibitem{Lein_2003}%
  \BibitemOpen
  \bibfield{author}{%
  \bibinfo {author} {\bibfnamefont{M.}~\bibnamefont{Lein}},\ }%
  \bibfield{journal}{%
  \bibinfo {journal} {J. Phys. B}\ }%
  \textbf{\bibinfo {volume} {36}},\ \bibinfo {pages} {L155} (\bibinfo {year}
  {2003})%
  \bibAnnoteFile{NoStop}{Lein_2003}%
\bibitem{Kjeldsen_2003}%
  \BibitemOpen
  \bibfield{author}{%
  \bibinfo {author} {\bibfnamefont{T.~K.}\ \bibnamefont{Kjeldsen}}, \bibinfo
  {author} {\bibfnamefont{C.~Z.}\ \bibnamefont{Bisgaard}}, \bibinfo {author}
  {\bibfnamefont{L.~B.}\ \bibnamefont{Madsen}},\ and\ \bibinfo {author}
  {\bibfnamefont{H.}~\bibnamefont{Stapelfeldt}},\ }%
  \bibfield{journal}{%
  \bibinfo {journal} {Phys. Rev. A}\ }%
  \textbf{\bibinfo {volume} {68}},\ \bibinfo {pages} {063407} (\bibinfo {year}
  {2003})%
  \bibAnnoteFile{NoStop}{Kjeldsen_2003}%
\bibitem{Fischer_2006}%
  \BibitemOpen
  \bibfield{author}{%
  \bibinfo {author} {\bibfnamefont{R.}~\bibnamefont{Fischer}}, \bibinfo
  {author} {\bibfnamefont{M.}~\bibnamefont{Lein}},\ and\ \bibinfo {author}
  {\bibfnamefont{C.~H.}\ \bibnamefont{Keitel}},\ }%
  \bibfield{journal}{%
  \bibinfo {journal} {Phys. Rev. Lett.}\ }%
  \textbf{\bibinfo {volume} {97}},\ \bibinfo {pages} {143901} (\bibinfo {year}
  {2006})%
  \bibAnnoteFile{NoStop}{Fischer_2006}%
\bibitem{McNaught_1997}%
  \BibitemOpen
  \bibfield{author}{%
  \bibinfo {author} {\bibfnamefont{S.~J.}\ \bibnamefont{McNaught}}, \bibinfo
  {author} {\bibfnamefont{J.~P.}\ \bibnamefont{Knauer}},\ and\ \bibinfo
  {author} {\bibfnamefont{D.~D.}\ \bibnamefont{Meyerhofer}},\ }%
  \bibfield{journal}{%
  \bibinfo {journal} {Phys. Rev. Lett.}\ }%
  \textbf{\bibinfo {volume} {78}},\ \bibinfo {pages} {626} (\bibinfo {year}
  {1997})%
  \bibAnnoteFile{NoStop}{McNaught_1997}%
\bibitem{Moore_1999}%
  \BibitemOpen
  \bibfield{author}{%
  \bibinfo {author} {\bibfnamefont{C.~I.}\ \bibnamefont{Moore}}, \bibinfo
  {author} {\bibfnamefont{A.}~\bibnamefont{Ting}}, \bibinfo {author}
  {\bibfnamefont{S.~J.}\ \bibnamefont{McNaught}}, \bibinfo {author}
  {\bibfnamefont{J.}~\bibnamefont{Qiu}}, \bibinfo {author}
  {\bibfnamefont{H.~R.}\ \bibnamefont{Burris}},\ and\ \bibinfo {author}
  {\bibfnamefont{P.}~\bibnamefont{Sprangle}},\ }%
  \bibfield{journal}{%
  \bibinfo {journal} {Phys. Rev. Lett.}\ }%
  \textbf{\bibinfo {volume} {82}},\ \bibinfo {pages} {1688} (\bibinfo {year}
  {1999})%
  \bibAnnoteFile{NoStop}{Moore_1999}%
\bibitem{Yamakawa_2004}%
  \BibitemOpen
  \bibfield{author}{%
  \bibinfo {author} {\bibfnamefont{K.}~\bibnamefont{Yamakawa}}, \bibinfo
  {author} {\bibfnamefont{Y.}~\bibnamefont{Akahane}}, \bibinfo {author}
  {\bibfnamefont{Y.}~\bibnamefont{Fukuda}}, \bibinfo {author}
  {\bibfnamefont{M.}~\bibnamefont{Aoyama}}, \bibinfo {author}
  {\bibfnamefont{N.}~\bibnamefont{Inoue}}, \bibinfo {author}
  {\bibfnamefont{H.}~\bibnamefont{Ueda}},\ and\ \bibinfo {author}
  {\bibfnamefont{T.}~\bibnamefont{Utsumi}},\ }%
  \bibfield{journal}{%
  \bibinfo {journal} {Phys. Rev. Lett.}\ }%
  \textbf{\bibinfo {volume} {92}},\ \bibinfo {pages} {123001} (\bibinfo {year}
  {2004})%
  \bibAnnoteFile{NoStop}{Yamakawa_2004}%
\bibitem{Gubbini_2005}%
  \BibitemOpen
  \bibfield{author}{%
  \bibinfo {author} {\bibfnamefont{E.}~\bibnamefont{Gubbini}}, \bibinfo
  {author} {\bibfnamefont{U.}~\bibnamefont{Eichmann}}, \bibinfo {author}
  {\bibfnamefont{M.}~\bibnamefont{Kalashnikov}},\ and\ \bibinfo {author}
  {\bibfnamefont{W.}~\bibnamefont{Sandner}},\ }%
  \bibfield{journal}{%
  \bibinfo {journal} {Phys. Rev. Lett.}\ }%
  \textbf{\bibinfo {volume} {94}},\ \bibinfo {pages} {053602} (\bibinfo {year}
  {2005})%
  \bibAnnoteFile{NoStop}{Gubbini_2005}%
\bibitem{Taieb_2001}%
  \BibitemOpen
  \bibfield{author}{%
  \bibinfo {author} {\bibfnamefont{R.}~\bibnamefont{Ta\"ieb}}, \bibinfo
  {author} {\bibfnamefont{V.}~\bibnamefont{V\'eniard}},\ and\ \bibinfo {author}
  {\bibfnamefont{A.}~\bibnamefont{Maquet}},\ }%
  \bibfield{journal}{%
  \bibinfo {journal} {Phys. Rev. Lett.}\ }%
  \textbf{\bibinfo {volume} {87}},\ \bibinfo {pages} {053002} (\bibinfo {year}
  {2001})%
  \bibAnnoteFile{NoStop}{Taieb_2001}%
\bibitem{Barth_2011}%
  \BibitemOpen
  \bibfield{author}{%
  \bibinfo {author} {\bibfnamefont{I.}~\bibnamefont{Barth}}\ and\ \bibinfo
  {author} {\bibfnamefont{O.}~\bibnamefont{Smirnova}},\ }%
  \bibfield{journal}{%
  \bibinfo {journal} {Phys. Rev. A}\ }%
  \textbf{\bibinfo {volume} {84}},\ \bibinfo {pages} {063415} (\bibinfo {year}
  {2011})%
  \bibAnnoteFile{NoStop}{Barth_2011}%
\bibitem{Barth_2013a}%
  \BibitemOpen
  \bibfield{author}{%
  \bibinfo {author} {\bibfnamefont{I.}~\bibnamefont{Barth}}\ and\ \bibinfo
  {author} {\bibfnamefont{O.}~\bibnamefont{Smirnova}},\ }%
  \bibfield{journal}{%
  \bibinfo {journal} {Phys. Rev. A}\ }%
  \textbf{\bibinfo {volume} {87}},\ \bibinfo {pages} {013433} (\bibinfo {year}
  {2013})%
  \bibAnnoteFile{NoStop}{Barth_2013a}%
\bibitem{Herath_2012}%
  \BibitemOpen
  \bibfield{author}{%
  \bibinfo {author} {\bibfnamefont{T.}~\bibnamefont{Herath}}, \bibinfo {author}
  {\bibfnamefont{L.}~\bibnamefont{Yan}}, \bibinfo {author}
  {\bibfnamefont{S.~K.}\ \bibnamefont{Lee}},\ and\ \bibinfo {author}
  {\bibfnamefont{W.}~\bibnamefont{Li}},\ }%
  \bibfield{journal}{%
  \bibinfo {journal} {Phys. Rev. Lett.}\ }%
  \textbf{\bibinfo {volume} {109}},\ \bibinfo {pages} {043004} (\bibinfo
  {month} {Jul}\ \bibinfo {year} {2012})%
  \bibAnnoteFile{NoStop}{Herath_2012}%
\bibitem{Keldysh_1965}%
  \BibitemOpen
  \bibfield{author}{%
  \bibinfo {author} {\bibfnamefont{L.~V.}\ \bibnamefont{Keldysh}},\ }%
  \bibfield{journal}{%
  \bibinfo {journal} {Zh. Eksp. Teor. Fiz.}\ }%
  \textbf{\bibinfo {volume} {47}},\ \bibinfo {pages} {1945} (\bibinfo {year}
  {1964})%
  \bibAnnoteFile{NoStop}{Keldysh_1965}%
\bibitem{Barth_2013b}%
  \BibitemOpen
  \bibfield{author}{%
  \bibinfo {author} {\bibfnamefont{I.}~\bibnamefont{Barth}}\ and\ \bibinfo
  {author} {\bibfnamefont{O.}~\bibnamefont{Smirnova}},\ }%
  \bibfield{journal}{%
  \bibinfo {journal} {Phys. Rev. A}\ }%
  \textbf{\bibinfo {volume} {88}},\ \bibinfo {pages} {013401} (\bibinfo {year}
  {2013})%
  \bibAnnoteFile{NoStop}{Barth_2013b}%
\bibitem{Klaiber_2013a}%
  \BibitemOpen
  \bibfield{author}{%
  \bibinfo {author} {\bibfnamefont{M.}~\bibnamefont{Klaiber}}, \bibinfo
  {author} {\bibfnamefont{E.}~\bibnamefont{Yakaboylu}},\ and\ \bibinfo {author}
  {\bibfnamefont{K.~Z.}\ \bibnamefont{Hatsagortsyan}},\ }%
  \bibfield{journal}{%
  \bibinfo {journal} {Phys. Rev. A}\ }%
  \textbf{\bibinfo {volume} {87}},\ \bibinfo {pages} {023417} (\bibinfo {year}
  {2013})%
  \bibAnnoteFile{NoStop}{Klaiber_2013a}%
\bibitem{Klaiber_2013b}%
  \BibitemOpen
  \bibfield{author}{%
  \bibinfo {author} {\bibfnamefont{M.}~\bibnamefont{Klaiber}}, \bibinfo
  {author} {\bibfnamefont{E.}~\bibnamefont{Yakaboylu}},\ and\ \bibinfo {author}
  {\bibfnamefont{K.~Z.}\ \bibnamefont{Hatsagortsyan}},\ }%
  \bibfield{journal}{%
  \bibinfo {journal} {Phys. Rev. A}\ }%
  \textbf{\bibinfo {volume} {87}},\ \bibinfo {pages} {023418} (\bibinfo {year}
  {2013})%
  \bibAnnoteFile{NoStop}{Klaiber_2013b}%
\bibitem{Bethe_1957}%
  \BibitemOpen
  \bibfield{author}{%
  \bibinfo {author} {\bibnamefont{{H. A. Bethe and E. E. Salpeter}}},\ }%
  \emph{\bibinfo {title} {Quantum mechanics of one- and two-electron atoms}}\
  (\bibinfo {publisher} {Academic Press, New York},\ \bibinfo {year} {1957})%
  \bibAnnoteFile{NoStop}{Bethe_1957}%
\end{thebibliography}%

\end{document}